\documentclass{article}
\usepackage{graphics}
\usepackage{epsfig}
\usepackage{cite}

\begin{document}

%\volnumpagesyear{0}{0}{000--000}{2005}
%\dates{received date}{revised date}{accepted date}

\title{Delay Induced Oscillations in a Fundamental Power System Model}

\author{Rajesh G. Kavasseri \\
Department of Electrical and Computer Engineering \\
North Dakota State University, Fargo, ND 58105 - 5285 \\email :
rajesh.kavasseri@ndsu.edu}

\date{}
\maketitle

\begin{abstract}In this paper, we study the dynamics and stability of a
fundamental power system model when a time delay is imposed on the
excitation of the generator. It is observed that sustained
oscillations can arise in an otherwise stable power system through a
delay induced Andronov-Hopf bifurcation. Numerical simulations are
conducted to explore the dynamics of the time delayed system after
the bifurcation which indicate period doublings culminating in a
strange attractor. \end{abstract}

\section{Introduction}
Complex nonlinear phenomena in power system models have been
extensively studied in the past. Arnold diffusion, \cite{salam}
and chaotic motions, \cite{kopell} have been studied in power
system models described by the swing equations. The classic period
doubling bifurcation route to chaos has been studied in a three
node power system model, \cite{dobson} using simplified models for
the generator \cite{abed},\cite{tan},\cite{chiang},\cite{wang} and
later, with detailed generator models in \cite{kavasseri_1}.
Hard-limit induced chaotic behavior in a single machine infinite
bus formulation was studied in \cite{mani_1} by extensive
numerical simulations where it was argued that the interaction of
hard limits and system transients led to sustained chaotic
behavior. In \cite{kavasseri_2}, hard limits were approximated by
a smooth function to study bifurcations in the three node power
system as
an extension of \cite{kavasseri_1}.\\

\noindent While considerable attention ([1]-[10]) has been devoted
to the study of power system models in the absence of time delays,
not as much attention has been paid to understand power system
dynamics in the presence of time delays with the exception of
\cite{martin}. In modern power systems with digital control
schemes, the acquisition, filtering and processing of signals can
lead to possible time delays, \cite{martin}. In \cite{martin}, the
authors demonstrated an effective control technique using
prediction to compensate for the delay in generator excitation. In
the deregulated power industry, communication delays arising from
the use of phasor measurement units in wide area measurement
systems could contribute to time delays in the acquisition of
control signals. Thus, the focus of this paper is to study the
effect of a time delay on the dynamics of a representative
power system model. \\

\begin{figure}[htbp]
\begin{center}
\includegraphics[height=2in,width=3in]{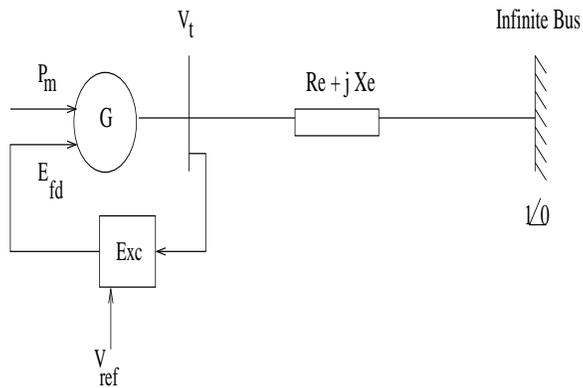}
\caption{Single Machine Infinite Bus System}\label{smib}
\end{center}
\end{figure}

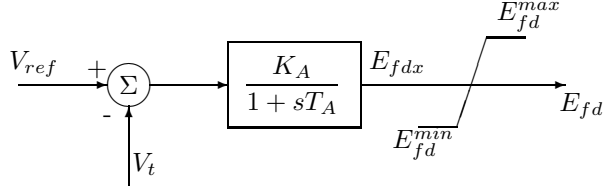
\begin{figure}[htbp]
\centerline{\unitlength=0.70mm
\special{em:linewidth 0.4pt}
\linethickness{0.4pt}
\begin{picture}(133.00,45.00)(15,15)
\put(47.00,33.00){\circle{8.25}}
\put(26.00,33.00){\vector(1,0){17.00}}
\put(47.00,14.00){\vector(0,1){15.00}}
\put(51.00,33.00){\vector(1,0){15.00}}
\put(66.00,25.00){\framebox(25.00,16.00)[cc]{}}
\put(98.00,37.00){\makebox(0,0)[cc]{$E_{fdx}$}}
\put(47.00,33.00){\makebox(0,0)[cc]{$\Sigma$}}
\put(29.00,37.00){\makebox(0,0)[cc]{$V_{ref}$}}
\put(41.00,36.00){\makebox(0,0)[cc]{+}}
\put(43.00,27.00){\makebox(0,0)[cc]{-}}
\put(50.00,18.00){\makebox(0,0)[cc]{$V_t$}}
\put(78.00,33.00){\makebox(0,0)[cc]{$\displaystyle{\frac {K_A}{1+sT_A}}$}}
\put(123.00,46.00){\makebox(0,0)[cc]{$E_{fd}^{max}$}}
\put(103.00,22.00){\makebox(0,0)[cc]{$E_{fd}^{min}$}}
\put(133.00,29.00){\makebox(0,0)[cc]{$E_{fd}$}}
\put(91.00,33.00){\vector(1,0){39.00}}
\put(115.00,42.00){\line(-1,-3){5.67}}
\put(109.33,25.00){\line(-1,0){7.33}}
\put(122.33,42.00){\line(-1,0){7.33}}
\end{picture}} \caption{Excitation Control System}
\label{exci}
\end{figure}

\noindent A Single Machine Infinite Bus (SMIB) system consisting of
a generator interfaced to the grid through a transmission line,
Figure \ref{smib} is considered in this study. The SMIB system
(\cite{anderson},\cite{padiyar}) has served as a long standing
paradigm to understand the dynamics of a single generator when
connected to a large power system. The simplicity of the SMIB system
facilitates analysis and thereby enables one to gain insight in to
the dynamic behavior of a synchronous generator. The
electro-mechanical equations of the generator are described by the
standard flux-decay model and the excitation is represented by a
single time constant system, Figure \ref{exci}. The excitation
voltage $E_{fd}$ is subject to a time delay $\tau$. The dynamics of
the resultant system is then described by delay differential
equations (DDEs). The reader is referred to \cite{hale},
\cite{bellman} for the general theory of DDE systems. In power
system models described by ODEs, Hopf bifurcations usually arise
when the loading of the generator, or control gains exceed certain
critical values, \cite{padiyar}. However as remarked earlier, little
attention (with the exception of \cite{martin}) has been paid to
study power system models with delays. Therefore, we focus on
understanding the role of a time delay on the dynamics of a
synchronous generator. Later on, we show that even if the control
gain and generator loadings are chosen to obtain a stable operating
point, a critical delay in the excitation can cause the normally
stable equilibrium to become unstable.

%The rest of this paper is organized as follows. \\

%\noindent Section 2 describes the system modeling. In Section 3,
%the analysis of delay induced Hopf bifurcation is carried out.
%Section 4 presents the results of numerical simulations and the
%conclusions are summarized in Section 5.

\section{System Model}
The system shown in Figure \ref{smib} illustrates a standard
single machine infinite bus (SMIB) system equipped with a fast
acting static excitation system (indicated by `Exc'). Assuming a
single time constant model for the excitation system shown in
Figure \ref{exci} and the flux decay model for the generator, the
dynamics of the single machine system in the presence of a time
delay in excitation ($E_{fd}$) can be described by, \\

\noindent  {\large $\Sigma_{\tau}$ }:

\begin{eqnarray}
\dot{\delta} = \omega_b \overline{\omega} \label{sys1}\\
\dot{\overline{\omega}}  = \frac{(P_m - P_e)}{2H} \\
\dot{E^{'}_q} = \frac{-E_q + (x_d - x^{'}_d)i_d + E_{fd}(t - \tau)}{T^{'}_{do}} \label{efd} \\
\dot{E_{fdx}} = \frac{-E_{fdx} + K_A(V_{ref} - V_t)}{T_A}
\label{sys4}
\end{eqnarray}

\begin{eqnarray}
 E_{fd} = \left \{ \begin{array}{ll}
                 E_{fdx} & \mbox{if $E_{fd}^{min}  \leq E_{fdx} \leq E_{fd}^{max} \  $} \\
                 E_{fd}^{min} & \mbox{if $E_{fdx} \leq E_{fd}^{min} \  $} \\
                 E_{fd}^{max} & \mbox{if $E_{fdx} \geq E_{fd}^{max} \ $} \end{array}
                 \right. \label{limi}
\end{eqnarray}

\noindent (\ref{limi}), describes a limiter which imposes upper
($E_{fd}^{max}$) and lower ($E_{fd}^{min}$) limits on the
excitation voltage $E_{fd}$ of the synchronous machine. While the
effect of excitation limits on the system dynamics and
bifurcations has been previously studied in \cite{mani_1} and
\cite{kavasseri_2}, the aim of the present paper is to understand
the role of the delay term $\tau$. Therefore, in the rest of the
analysis which demands smoothness of the models, the excitation
limits are ignored which implies that $E_{fdx} = E_{fd}$. The rest
of the variables $(v_d, i_d, v_q,i_q$) are given by,

\begin{eqnarray}
i_d = \frac{\cos(\delta) - E^{'}_q}{x^{'}_d + x_e},\;\; i_q = \frac{\sin(\delta)}{x_q + x_e} \\
v_d = -\frac{x_q \sin(\delta)}{x_q + x_e}, \;\; v_q =
\frac{x^{'}_d\cos(\delta) - E^{'}_q}{x^{'}_d + x_e}
\end{eqnarray}

\noindent The electrical power $P_e$ and terminal voltage $V_t$
are expressed in terms of the state variables from,
\begin{equation}
P_e= v_d i_d + v_q i_q,\;\; V_t = \sqrt{v^2_d + v^2_q}
\end{equation}

\section{Local Stability}
The local asymptotic stability of the  equilibrium point for the
system can be studied from the roots of the characteristic
polynomial corresponding to the linearisation of $\Sigma_{\tau}$
given by

\begin{equation}
|\lambda I - A | = 0 \label{chpolyA}
\end{equation}

where,

\begin{equation} A = \left [  \begin{array}{cccc}
0 & \omega_b  & 0 & 0\\
-\frac{K_1}{2H} &  0 & -\frac{K_2}{2H} & 0 \\
-\frac{K_4}{T^{'}_{do}} & 0 & -\frac{1}{K_3T^{'}_{do}} &
e^{-\lambda \tau}/{T^{'}_{do}} \\
-\frac{K_AK_5}{T_A} & 0 & -\frac{K_A K_6}{T_A} & -\frac{1}{T_A}
\end{array} \right ] = 0 \label{systA}
\end{equation}

\noindent In the linearisation of a single machine system
expressed in  (\ref{systA}), the parameters $K_1, \dots K_6$ are
known as Heffron-Phillips constants, \cite{anderson},
\cite{padiyar} which depend on the equilibrium and the machine
parameters (see Appendix A). Substituting for $A$ from
(\ref{systA}) in to (\ref{chpolyA}) yields the characteristic
polynomial for the system which can be expressed as

\begin{equation}
f_{\tau}(\lambda) = f_1(\lambda) +  e^{-\lambda \tau} g_1(\lambda)
= 0 \label{charpoly}
\end{equation}

\noindent where,

\begin{eqnarray}
f_1(\lambda) = \lambda^4 + \lambda^3 \beta_1 + \lambda^2 \beta_2 +
\lambda \beta_3 + \beta_4 \label{f1} \\
g_1(\lambda) = \lambda^2 \beta_5 + \beta_6 \label{g1}
\end{eqnarray}

\noindent The coefficients $\beta_1, \dots \beta_6$ are given by,
\begin{eqnarray}
\beta_1 = K^{\prime}_3 + 1/T_A \\
\beta_2 = K^{\prime}_3/T_A + K^{\prime}_1\omega_b +
K^{\prime}_6/T^{'}_{do} \\
\beta_3 = \omega_b(K^{\prime}_1K^{\prime}_3 -
K^{\prime}_2K^{\prime}_4) + K^{\prime}_1 \omega_b/T_A \\
\beta_4 =\omega_b(K^{\prime}_1K^{\prime}_3 -
K^{\prime}_2K^{\prime}_4) \\
\beta_5 = K^{'}_6 T^{'}_{do} \\
\beta_6 = \omega_b(K^{\prime}_1K^{\prime}_6-
K^{\prime}_2K^{\prime}_5)/T^{'}_{do}
\end{eqnarray}

\noindent with \begin{eqnarray} K^{\prime}_1 = K_1/2H, \;
K^{\prime}_2 = K_2/2H,\;K^{\prime}_3 = 1/K_3 T^{'}_{do}, \nonumber
\\ \;K^{\prime}_4 = K_4/T^{'}_{do},\;K^{\prime}_5= K_AK_5/T_A,\;
K^{\prime}_6 = K_AK_6/T_A \nonumber \end{eqnarray}

\noindent In the absence of delay, by setting $\tau=0$ in
(\ref{charpoly}) we note that the characteristic polynomial is
described by $f_0 (\lambda) = f_1(\lambda) + g_1(\lambda) = 0$. In
the present study, the control gain, machine loading and other
parameters are chosen so that the equilibrium point of the system
without delay is locally asymptotically stable. With this
assumption, the coefficients $\beta_i, i = 1 \dots 6$ are all
positive for typical values of machine parameters and loading
conditions. In the following section, we shall study the local
stability of the equilibrium as the delay $\tau$ is varied.

\section{Delay Induced Oscillations}
Local stability of the equilibrium point of $\Sigma_{\tau}$ is
governed by the roots of the characteristic polynomial
$f_{\tau}(\lambda) = 0$, (\ref{charpoly}). Continuous dependence
of the roots of (\ref{charpoly}) on $\tau$ implies that there
exists $\tau_{c}$ such that $\Re e \{{\lambda}\} < 0$ when $\tau
\in [0, \tau_{c})$. A loss of asymptotic stability of the
equilibrium occurs when $\Re e \{{\lambda}\} = 0$ for some
critical value of $\tau = \tau_{c}
> 0$. Suppose the characteristic polynomial (\ref{charpoly}) has a
pair of purely imaginary roots $\lambda = \pm i\omega_{c}$ when
$\tau = \tau_{c}$, we get

\begin{equation}
f_1(i \omega) + e^{-i \omega \tau} g_1 (i \omega) = 0
\end{equation}

\noindent Substituting for $f_1(.) \; $and $ g_1(.)$ from
(\ref{f1}) and (\ref{g1}) we get,

\begin{eqnarray}
\beta_c + \beta_a \cos (\omega \tau) = 0  \label{sol1}\\
\beta_b - \beta_a \sin (\omega \tau) = 0 \label{sol2}
\end{eqnarray}

\noindent where,

\begin{eqnarray}
\beta_a = \beta_6 - \omega^2 \beta_5 \\
\beta_b = \omega \beta_3 - \beta_1 \omega^3 \\
\beta_c = \omega^4 - \omega^2 \beta_2 + \beta_4
\end{eqnarray}

\noindent From (\ref{sol1}) and (\ref{sol2}), the pair
($\omega_{c}, \tau_{c}$) for which the system (\ref{charpoly}) has
a pair of purely imaginary roots can be found from

\begin{eqnarray}
\beta_a^2 = \beta_b^2 + \beta_c^2  \label{solw}\\
\tan(\omega \tau) =  -\frac{\beta_b}{\beta_c} \label{soltau}
\end{eqnarray}

\noindent By implicit differentiation of (\ref{charpoly}), we get
\begin{equation}
\frac{d\lambda}{d \tau} = \frac{\lambda e^{-\lambda \tau}
g_1(\lambda)}{f^{\prime}_1(\lambda) + g^{\prime}_1(\lambda) - \tau
e^{-\lambda \tau} g_1(\lambda)} \label{impli}
\end{equation}

\noindent Setting $\lambda = i \omega_{c}$ and $\tau = \tau_{c}$
in (\ref{impli}), using (\ref{solw}), (\ref{soltau}) and
simplifying the terms yields

\begin{equation}
\sigma = \Re e \{\frac{d\lambda}{d \tau}\} |_{\tau_{c},\omega_{c}}
= \frac{2 \omega_{c} \beta_a (\beta_2 - \beta_5)}{D^2}
\label{sigma}
\end{equation}

\noindent where,

%\begin{eqnarray}

\begin{equation}
D^2  = D_a^2 + D_b^2 \end{equation}

\begin{equation} D_a  =   \beta_3 - 3 \beta_1\omega_c^2 - \beta_a \tau_c
\cos(\omega_c \tau_c) + 2\omega_c \beta_2 \sin(\omega_c \tau_c)
\end{equation} \begin{equation} D_b = 2 \omega_c \beta_2 - 4 \omega^3_c + \beta_a \tau_c
\sin(\omega_c \tau) + 2 \omega_c \beta_2 \cos(\omega_c \tau_c)
\end{equation}
%\end{eqnarray}

\noindent Therefore, the sign of $\sigma$ is governed by the terms
in the numerator of (\ref{sigma}). Substituting for the terms
$\beta_2, \beta_5$ from Appendix B we get,
\begin{equation} \beta_2  - \beta_5 = \frac{1 -
K_3T_eK_6}{T^{'}_{do}K_3 T_e} + \frac{K_1\omega_b}{2H}
\end{equation}

\noindent Under normal operating conditions for a power system,
the Heffron-Phillips constants satisfy, \begin{eqnarray} K_1 > 0
\\ 0 < K_3 < 1 \\ 0 < K_6 < 1
\end{eqnarray}

\noindent In addition, fast acting excitation systems typically
have a small time constant so that $0 < T_e < 1$. In view of the
above conditions, we note that

\begin{equation}
(\beta_2 -\beta_5) > 0
\end{equation}

\noindent The equations for $\beta_a(\omega_c) \ne 0$ turn out to be
complicated, making explicit analysis difficult. However numerical
solutions of (\ref{solw}) over realistic ranges of operating
conditions yield nonzero values for the term $\beta_a$ at the
critical point in which case, $\sigma > 0$. Therefore, the results
of the analysis so
far can be summarized in the following proposition. \\

\noindent {\bf Proposition 1 }: The system $\Sigma_{\tau}$
satisfies the sufficient conditions for a delay induced
Andronov-Hopf-bifurcation under the following assumptions.

\begin{itemize} \item [[A1]] the system without delay $\Sigma_0$ is stable $\Rightarrow
f_0(\lambda) = 0$ has no roots on the right half plane. \item
[[A2]] if $\tau \ne 0$, then $\lambda = 0$ is not a root of
$f_{\tau}(\lambda)$ (\ref{charpoly}).
\end{itemize}

\noindent The assumptions [A1] and [A2] indicate that at the
critical point, the characteristic polynomial has a purely
imaginary pair of roots and all other roots of the polynomial have
strictly negative real parts. The delay induced loss of operating
point stability leads to the emergence of periodic solutions. The
stability of the resulting periodic motion can be inferred from
the sign of a certain quantity which proves to be too difficult in
this case. Therefore, we resort to numerical simulations in the
following section to study the dynamics of the system after delay
induced instability.

\section{Simulation Results}
\begin{figure}[htbp]
\begin{center}
\includegraphics[height=2.5in,width=3.5in]{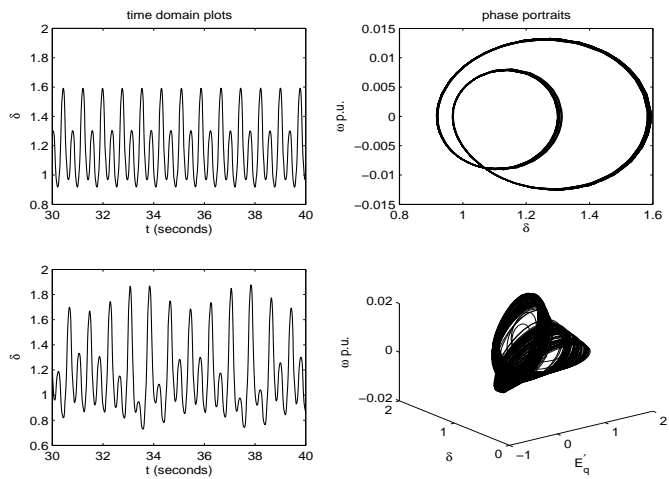}
\caption{Simulation of system trajectory beyond $\tau_c$. Upper
panel indicates a period doubling when $\tau = 0.059$ followed by
a strange attractor in the lower panel when $\tau =
0.061$.}\label{pla2}
\end{center}
\end{figure}

All numerical integrations were carried out using {\em dde23} in
{\em MATLAB}. A numerical simulation of the system $\Sigma_{\tau}$
is shown in Figure \ref{pla1} assuming typical values for the
machine loading and control gain $K_A$ (supplied in Appendix A)
that results in a stable operating point in the absence of delay.
Then the delay $\tau$ is increased gradually from zero up to the
critical value $\tau_{c}$ when the equilibrium point loses
stability. The rotor angle $\delta$ is plotted in all time domain
simulations. The transition from damped oscillations (when $0 <
\tau < \tau_{c}$) to sustained oscillations (when $\tau
> \tau_{c}$) can be noted from Figure \ref{pla1}. Incidentally, the
solution of (\ref{solw}) and (\ref{soltau}) yield $\tau_c =
0.0533$ which closely matches the critical delay observed by
simulation, i.e. $\tau_{cri} = 0.055$. When the delay $\tau$ is
increased further, a period doubling (upper panel of Figure
\ref{pla2}) is noticed which eventually culminates in a strange
attractor (lower panel of Figure \ref{pla2}).

\begin{figure}[htbp]
\begin{center}
\includegraphics[height=2.3in,width=3.5in]{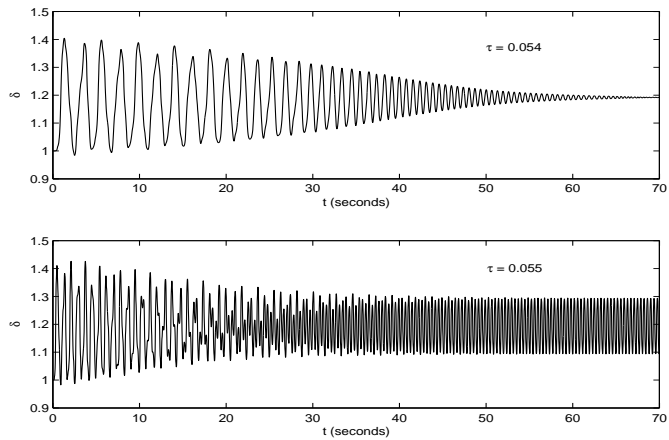}
\caption{The onset of oscillatory activity in the time delayed
SMIB system. The critical delay $\tau_c = 0.055$.  Upper panel
shows a stable trajectory when $\tau = 0.054 < \tau_{c}$. Lower
panel shows sustained oscillations when $\tau =
\tau_{c}$.}\label{pla1}
\end{center}
\end{figure}

\section{Conclusions}
Power system models are capable of displaying a rich variety of
nonlinear phenomena which have been formally studied by
bifurcation theory. In particular, the emergence of oscillations
in power system models has been attributed to Hopf bifurcations
which have been shown to arise when the control gains or machine
loading exceed critical values. In this paper, the dynamics of a
fundamental power system model is studied when the field voltage
of the generator is subject to a time delay. It is shown by
preliminary analysis that a nominally stable operating point can
be destabilized when the time delay exceeds a critical value
$\tau_{c}$ through a Andronov-Hopf bifurcation. The behavior of
the system for $\tau> \tau_{c}$ is further studied by numerical
simulations which indicate that the system undergoes a period
doubling which eventually culminates in a strange attractor.
Further analysis to determine the bifurcations of the ensuing
limit cycle would be an interesting subject for future research.

\section*{Appendix A}
The Heffron-Phillips constants for a lossless network are given by
\begin{eqnarray}
K_1 = \frac{E_{qo}\cos(\delta_o)}{(x_e + x_q)} + \frac{(x_q -
x_d^{\prime})i_{qo} \sin(\delta_o)}{x_e + x^{\prime}_d} \\
K_2 = \frac{\sin(\delta_o)}{x_e + x^{\prime}_d}, \;\; K_3 = \frac{x_e + x^{\prime}_d}{x_d + x_e} \\
K_4 = \frac{(x_d - x^{\prime}_d) \sin(\delta_o)}{x_e +
x^{\prime}_d}\\
K_5 = \frac{-x_q v_{do}\cos(\delta_o)}{(x_e + x_q)V_{to}} -
\frac{-x^{\prime}_d v_{qo}\sin(\delta_o)}{(x_e +
x^{\prime}_d)V_{to}} \\
K_6 = \frac{x_e v_{qo}}{V_{to}(x_e + x^{\prime}_d})
\end{eqnarray}
where the subscript `$o$' indicates the values of the variables at
the equilibrium point which can be solved by setting
(\ref{sys1})- (\ref{sys4}) to zero. \\

\section*{Appendix B}
\noindent System Parameters : \\
$x_d = 1.79, \;x_q = 1.71, \; x^{\prime}_d = 0.169,\; P_m = 0.9,\;
T^{\prime}_{do} = 4.3,\;H = 5, \; \omega_b = 377, \; x_e = 0.4,\;\; R_e = 0$.\\

\noindent Exciter Parameters : \\
$K_A = 200,\; T_A = 0.05$.

\end{document}